\documentclass{article}

\usepackage{fullpage}

\usepackage{enumitem}
\setlist{nosep, leftmargin=14pt}
 
\usepackage{mwe} %
\usepackage{bm}
\usepackage{amsfonts}
\usepackage{url}            %
\usepackage{booktabs}       %
\usepackage{amsfonts}       %
\usepackage{nicefrac}       %
\usepackage{microtype}     %
\usepackage{xcolor}         %
\usepackage{bm}         %
\usepackage{graphicx}  		%
\usepackage{amsmath} %
\usepackage{cite} %
\usepackage{diagbox}
\usepackage{enumitem}
\usepackage{siunitx}
\usepackage{multirow}
\usepackage{tabularx}
\usepackage{hhline}
\usepackage{arydshln}		%
\usepackage{xcolor}
\usepackage{mathtools} %
\usepackage{amsthm}			%
\usepackage{amssymb}			%
\usepackage{algorithmic}	%
\usepackage{algorithm}
\usepackage{pifont}%
\newcommand{\cmark}{\ding{51}}%
\newcommand{\xmark}{\ding{55}}%
\definecolor{lightgreen}{rgb}{.9,1,.9}
\definecolor{lightred}{rgb}{1,.415,.415}
\definecolor{lightblue}{rgb}{.415,.415,1}

\def\argmin{\mathop{\mathrm{arg\,min}}} %

\def\lim{\mathop{\mathrm{lim}}} %
 
\def\max{\mathop{\mathrm{max}}}

\def\log{\mathrm{log}}

\newcommand{\norm}[1]{\left\lVert#1\right\rVert}

\def\ebm{{\bm{e}}}

\def\xbm{{\bm{x}}}

\def\ybm{{\bm{y}}}
\def\zbm{{\bm{z}}}

\def\vbm{{\bm{v}}}

\def\thetabm{{\bm{\theta}}}

\def\Abm{{\bm{A}}}

\def\xbmhat{{\widehat{\bm{x}}}}

\def\R{\mathbb{R}}

\begin{document}

\title{Plug-and-Play Posterior Sampling for Blind Inverse Problems}
\date{}

\author{
    Anqi Li$^{*}$, Weijie Gan$^{*}$, and Ulugbek S.~Kamilov \\
\small Washington University in St.~Louis, MO, USA \\
$^{*}$\small These authors contributed equally.\\
\small\texttt{$\{$anqi.li1, weijie.gan, kamilov$\}$@wustl.edu}}

\maketitle

\begin{abstract}
We introduce \emph{Blind Plug-and-Play Diffusion Models (Blind-PnPDM)} as a novel framework for solving blind inverse problems where both the target image and the measurement operator are unknown. Unlike conventional methods that rely on explicit priors or separate parameter estimation, our approach performs posterior sampling by recasting the problem into an alternating Gaussian denoising scheme. We leverage two diffusion models as learned priors: one to capture the distribution of the target image and another to characterize the parameters of the measurement operator. This PnP integration of diffusion models ensures flexibility and ease of adaptation. Our experiments on blind image deblurring show that Blind-PnPDM outperforms state-of-the-art methods in terms of both quantitative metrics and visual fidelity. Our results highlight the effectiveness of treating blind inverse problems as a sequence of denoising subproblems while harnessing the expressive power of diffusion-based priors.
\end{abstract}

\section{Introduction}

Many problems in computational imaging, biomedical imaging, and computer vision can be formulated as \emph{inverse problems} involving the recovery of high-quality images from low-quality observations. Imaging inverse problems are generally ill-posed, which means that multiple plausible clean images could lead to the same observation. It is thus common to introduce prior models on the desired images. While the literature on prior modeling of images is vast, current methods are often based on \emph{deep learning (DL)}, where a deep model is trained to map observations to images~\cite{McCann.etal2017, Lucas.etal2018, Ongie.etal2020}. 

\emph{Plug-and-play (PnP) methods}~\cite{Venkatakrishnan.etal2013, Sreehari.etal2016} are widely-used DL frameworks for solving imaging inverse problems. PnP methods circumvent the need to explicitly describe the full probability density of images by specifying image priors using image denoisers. The integration of state-of-the-art deep denoisers with physical measurement models within PnP has been shown to be effective in a number of inverse problems, including image super-resolution, phase retrieval, microscopy, and medical imaging~\cite{Metzler.etal2018, Zhang.etal2017a, Meinhardt.etal2017, Dong.etal2019, Zhang.etal2019, Wei.etal2020, Zhang.etal2022, Liu.etal2022} (see also recent reviews~\cite{Ahmad.etal2020, Kamilov.etal2023}). Practical success of PnP has also motivated novel extensions, theoretical analyses, statistical interpretations~\cite{Chan.etal2016, Romano.etal2017, Buzzard.etal2017, Reehorst.Schniter2019, Sun.etal2018a, Sun.etal2019b, Ryu.etal2019, Xu.etal2020, Liu.etal2021b, Kadkhodaie.Simoncelli2021, Hurault.etal2022, hurault2022proximal, gan2024block}.

While traditional PnP methods produce only point estimates, there is growing interest in PnP-based sampling methods that can produce solutions by sampling from the posterior distribution~\cite{Laumont.etal2022, Coeurdoux.etal2024, wu2024principled}. In particular, \emph{Plug-and-play Diffusion Models (PnPDM)}~\cite{wu2024principled} has proposed to leverage pre-trained \emph{diffusion models (DMs)} as priors within PnP sampling formulation. Despite this progress, the existing work on PnP sampling has primarily focused on the problem of image recovery where the measurement operator is known exactly. There is little work on PnP sampling for \emph{blind} inverse problems, where both the image and the measurement operator are unknown. This form of blind inverse problems are ubiquitous in computational imaging with well-known applications such as blind deblurring~\cite{campisi2017blind} and parallel magnetic resonance imaging (MRI)~\cite{Fessler2020}. In this paper, we address this gap by developing a new PnP sampling approach, called \emph{Blind-PnPDM}, that uses pre-trained DMs as priors over both the unknown measurement model and the unknown image, and efficiently solves the joint estimation task. We show the practical relevance of Blind-PnPDM by solving joint estimation problems in blind deblurring. Our numerical results show the potential of DMs to act as PnP priors over the measurement operators as well as images. Our work thus addresses a gap in the current PnP literature by providing a new efficient and principled sampling framework applicable to a wide variety of blind imaging inverse problems.

\section{Method}
\subsection{Preliminaries}

Consider an inverse problem that aims to recover a clean image $\xbm \in \R^n$ from its noisy measurement $\ybm \in \R^m$, modeled by the linear system 
$
\ybm = \Abm(\thetabm)\xbm + \ebm,
$
where $\Abm(\thetabm) \in \R^{m \times n}$ is the measurement operator parameterized by $\thetabm$, and $\ebm \in \R^m$ represents the noise. When $\thetabm$ is assumed to be known, a common approach to solving inverse problems is to formulate an optimization problem:
\begin{equation}
    \label{equ:optimization}
    \xbmhat \in \argmin_{\xbm \in \R^n} f(\xbm) + g(\xbm)\ ,
\end{equation}
where $f$ is the data-fidelity term, which ensures consistency with the measurement $\ybm$, and $g$ is the regularizer that encodes prior knowledge of $\xbm$. In computational imaging, a widely used data-fidelity term is the least-squares function $f(\xbm) = \tfrac{1}{2}\norm{\Abm\xbm - \ybm}_2^2$, while total variation (TV) is a common choice for the regularizer. PnP methods are popular DL frameworks that incorporate state-of-the-art denoisers as implicit regularizers. For instance, the PnP variant of the proximal gradient method, known as PnP-ISTA, can be expressed as the fixed-point iteration:
$
\xbm^+ = D_\sigma(\xbm - \gamma \nabla f(\xbm)),
$
where $D_\sigma$ is a denoiser parameterized by $\sigma > 0$, which controls the denoising strength, and $\gamma > 0$ is the step size.

Traditional PnP methods yield only point estimates. Recently, there has been growing interest in PnP-based sampling methods that generate solutions by sampling from the posterior distribution 
$
p(\xbm|\ybm) \propto p(\ybm|\xbm)p(\xbm) = \exp(-f(\xbm;\ybm) - g(\xbm)),
$
particularly by leveraging diffusion models (DMs) as priors. Notable recent works include \emph{PnPDM}~\cite{wu2024principled} and \emph{Diffusion Plug-and-Play (DPnP)}~\cite{xu2024provably}. These methods leverage Split Gibbs Samplers (SGS)~\cite{vono2019split} to alternate between two sampling steps that separately involve the likelihood and prior, where the likelihood step can be addressed using conventional sampling techniques whereas the prior step connects to the unconditional image generation process within the general framework of DMs. To be specific, these methods adopt a variable-splitting strategy by introducing an auxiliary variable $\zbm$, resulting in the augmented distribution:
\begin{equation}
    \pi(\xbm,\zbm) \propto \exp\Big(-f(\zbm;\ybm) - g(\xbm) - \tfrac{1}{2\rho^2}\norm{\xbm - \zbm}_2^2\Big),
\end{equation}
where $\rho > 0$ is a hyperparameter controlling the coupling strength between $\xbm$ and $\zbm$. Starting from an initial estimate $\xbm^{(0)}$, for iterations $k = 0, \ldots, K$, these methods alternately sample between
\begin{subequations}
\begin{align}
    \zbm^{(k)} &\sim \pi^{Z|X=\xbm^{(k)}}(\zbm) \\
    &\propto \exp\Big(-f(\zbm;\ybm) - \tfrac{1}{2\rho^2}\norm{\xbm^{(k)} - \zbm}_2^2\Big)\ ,
\end{align}
\end{subequations}
and
\begin{subequations}
\label{equ:sampling-x-1}
\begin{align}
    \xbm^{(k+1)} &\sim \pi^{X|Z=\zbm^{(k)}}(\xbm) \\
    &\propto \exp\Big(-g(\xbm) - \tfrac{1}{2\rho^2}\norm{\xbm - \zbm^{(k)}}_2^2\Big).
\end{align}
\end{subequations}
The prior step \eqref{equ:sampling-x-1} essentially corresponds to denoising $\zbm$ with a noise level of $\rho$. The key idea is to relate this sampling step to the unconditional image generation capability of DMs. For instance, in PnPDM, when implementing DMs within the Elucidated Diffusion Model (EDM) framework~\cite{karras2022elucidating}, the prior update \eqref{equ:sampling-x-1} can be performed through the following steps: (1) find $t^*$ such that $\sigma(t^*) = \rho$, (2) initialize $\xbm_{t^*} = s(t^*) \zbm^{(k)}$, and (3) run the DM inference process from $t^*$ to $0$. Here, $\sigma(t)$ and $s(t)$ are predefined noise level and scaling schedules in EDM.

Despite these advancements, existing PnP sampling methods have predominantly focused on image recovery problems where the measurement operator is known exactly. In this paper, we aim to extend PnP sampling to the \emph{blind inverse problem} setting, where the measurement parameters $\thetabm$ are unknown.

\subsection{Blind-PnPDM}

\begin{algorithm}[t]
\caption{Blind Plug-and-Play Diffusion Models}
\label{alg}
\begin{algorithmic}[1]
\STATE \textbf{input}: initialization $\xbm^{(0)}\in\R^n$ and $\thetabm^{(0)}\in\R^b$, total number of iteration $K>0$, coupling strength schedule $\{\rho_x^{(k)} > 0\}_{k=0}^{K-1}$ and $\{\rho_\theta^{(k)} > 0\}_{k=0}^{K-1}$, pre-trained diffusion priors $D_\alpha$ and $D_\beta$ for images and parameters of forward model, respectively.

\FOR{$k = 0, ..., K-1$}

\STATE {\color{gray} \# Sample from $p(\xbm|\ybm,\thetabm^{(k)})$}
\STATE $\zbm^{(k)} = \mathsf{LikelihoodStep}_{\xbm}(\xbm^{(k)}, \thetabm^{(k)}, \rho_x^{(k)})$
\STATE $\xbm^{(k+1)} = \mathsf{PriorStep}_{\xbm}(\zbm^{(k)}, \rho_x^{(k)}, D_\alpha)$
\STATE
\STATE {\color{gray} \# Sample from $p(\thetabm|\ybm,\xbm^{(k+1)})$}
\STATE $\vbm^{(k)} = \mathsf{LikelihoodStep}_{\thetabm}(\xbm^{(k+1)}, \thetabm^{(k)}, \rho_\theta^{(k)})$
\STATE $\thetabm^{(k+1)} = \mathsf{PriorStep}_{\thetabm}(\vbm^{(k)}, \rho_\theta^{(k)}, D_\beta)$
\ENDFOR
\end{algorithmic}
\end{algorithm}

\begin{figure*}
    \centering
    \includegraphics[width=\linewidth]{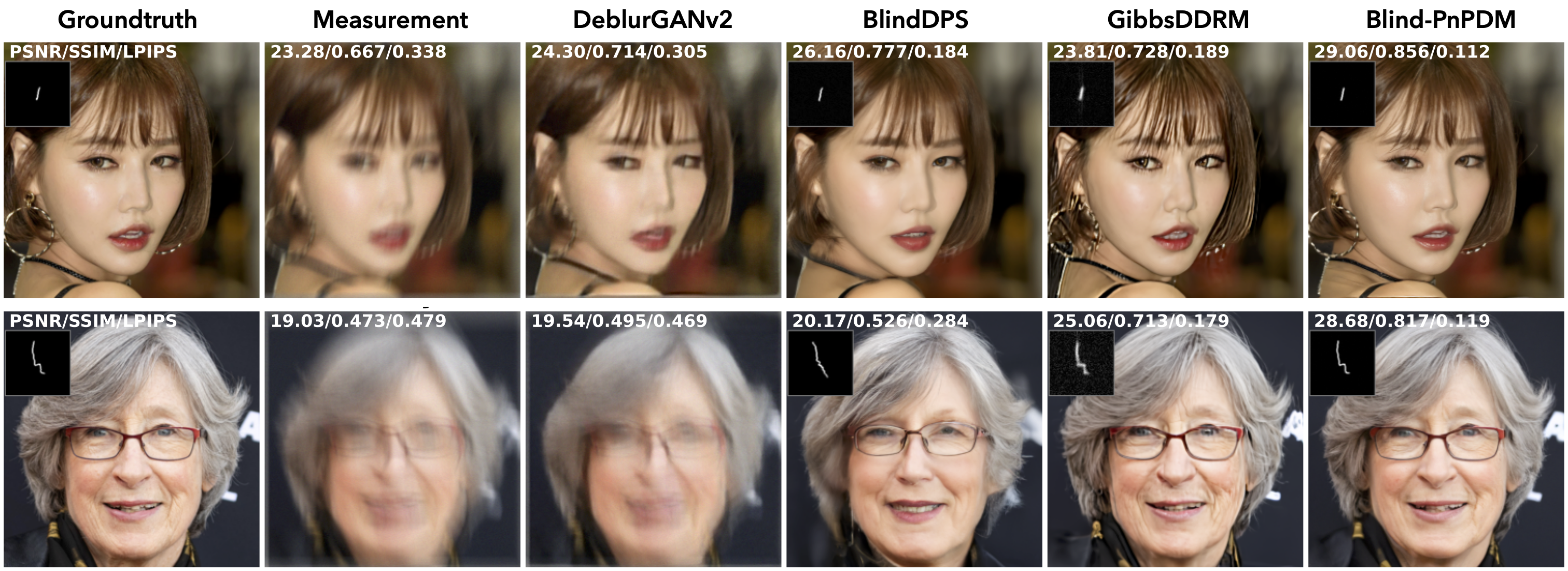}
    \caption{Illustration of the results obtained from several well-known methods for blind image deblurring with a motion kernel. The squares at the top of each image display the estimated kernels. The values in the top-left corner of each image indicate the PSNR, SSIM, and LPIPS metrics for the corresponding method. This figure demonstrates that Blind-PnPDM can reconstruct both the image and the kernel with finer details and greater consistency with the ground truth compared to other baseline methods.}
    \label{fig:motion}
\end{figure*}

\begin{figure*}
    \centering
    \includegraphics[width=\linewidth]{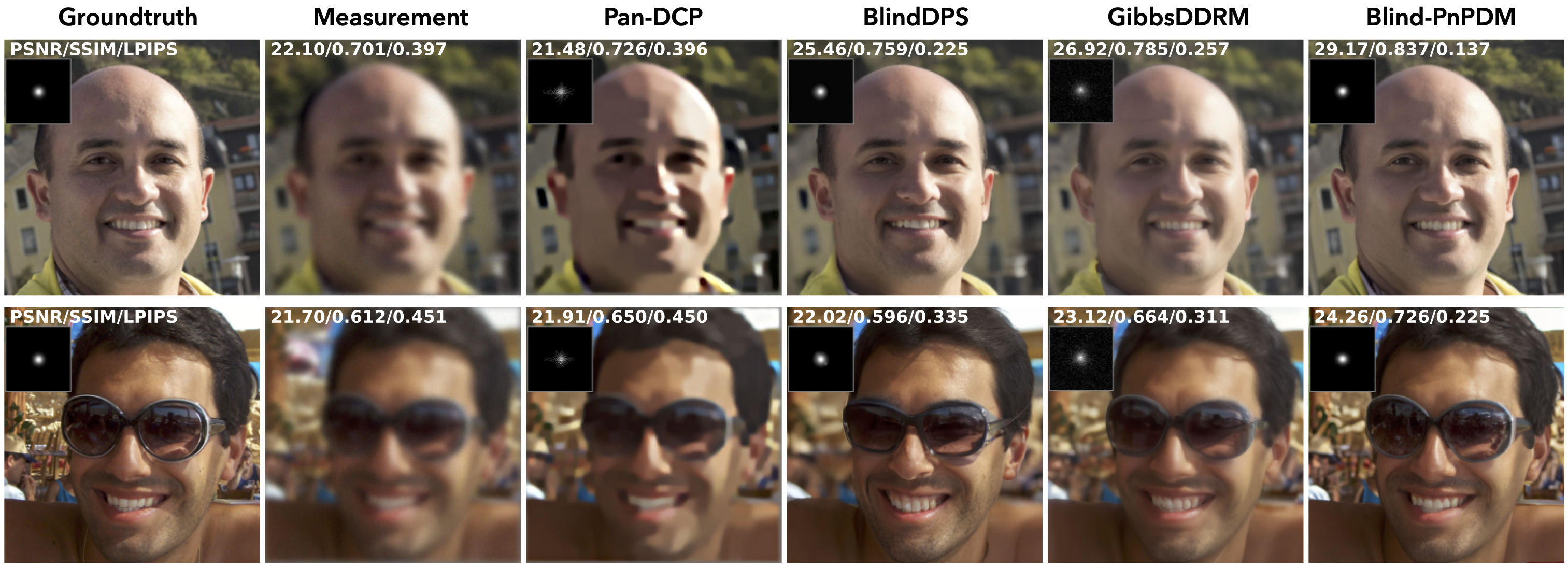}
    \caption{Illustration of the results obtained from several well-known methods for blind image deblurring with a Gaussian kernel. The squares at the top of each image display the estimated kernels. The values in the top-left corner of each image indicate the PSNR, SSIM, and LPIPS metrics for the corresponding method. Notably, Blind-PnPDM produces background details with finer textures, whereas other methods yield smoother results.}
    \label{fig:gaussian}
\end{figure*}

Our goal is to sample from the joint posterior distribution:
\begin{equation}
\begin{aligned}
    p(\xbm,\thetabm|\ybm) & \propto p(\ybm|\xbm,\thetabm)p(\xbm)p(\thetabm) \\
    & = \exp\big(-f(\xbm,\thetabm;\ybm) - g(\xbm) - h(\thetabm)\big),
\end{aligned}
\end{equation}
where $f(\xbm,\thetabm;\ybm) \coloneqq -\log p(\ybm|\xbm,\thetabm)$, $g(\xbm) \coloneqq -\log p(\xbm)$, and $h(\thetabm) \coloneqq -\log p(\thetabm)$ are the potential functions corresponding to the likelihood, the prior on the image, and the prior on the forward model parameters, respectively.
The complete pseudocode of Blind-PnPDM is presented in Algorithm~\ref{alg}. Instead of directly sampling from the joint posterior, Blind-PnPDM leverages Gibbs sampling to alternately draw samples from the $\xbm$- and $\thetabm$-conditional distributions. Specifically, starting from initial values $\xbm^{(0)}$ and $\thetabm^{(0)}$, for iterations $k = 0, \ldots, K-1$, we alternately sample:
\begin{subequations}
\label{equ:sampling-x}
   \begin{align} 
    \xbm^{(k+1)} &\sim p(\xbm|\ybm,\thetabm^{(k)}) \\
    &= \exp\big(-f(\xbm;\thetabm^{(k)},\ybm) - g(\xbm)\big),
    \end{align} 
\end{subequations}
and
\begin{subequations}
\label{equ:sampling-theta}
\begin{align}
    \thetabm^{(k+1)} &\sim p(\thetabm|\ybm,\xbm^{(k+1)}) \\
    &= \exp\big(-f(\thetabm;\xbm^{(k+1)},\ybm) - h(\thetabm)\big).
\end{align}
\end{subequations}
Inspired by PnPDM~\cite{wu2024principled}, we decompose the sampling of \eqref{equ:sampling-x} into two steps: $\mathsf{LikelihoodStep}_\xbm$ and $\mathsf{PriorStep}_\xbm$, following the SGS framework. Similarly, \eqref{equ:sampling-theta} is decomposed into $\mathsf{LikelihoodStep}_\thetabm$ and $\mathsf{PriorStep}_\thetabm$.
In the $\mathsf{LikelihoodStep}_\xbm$ at iteration $k$, we sample:
\begin{equation*}
    \zbm^{(k)} \sim \exp\Big(-\tfrac{1}{2}\norm{\ybm - \Abm(\thetabm^{(k)})\ \zbm}_2^2 - \tfrac{1}{2\rho_x^2}\norm{\xbm^{(k)} - \zbm}_2^2\Big),
\end{equation*}
where $\zbm \in \R^n$ is an auxiliary variable for the image, and $\rho_x > 0$ is a hyperparameter controlling the coupling strength between $\xbm$ and $\zbm$.
In the $\mathsf{PriorStep}_\xbm$, we sample:
\begin{equation}
    \xbm^{(k+1)} \sim \exp\Big(-g(\xbm) - \tfrac{1}{2\rho_x^2}\norm{\xbm - \zbm^{(k)}}_2^2\Big).
\end{equation}
This step is implemented using a DM $D_\alpha$ within the EDM framework~\cite{karras2022elucidating} as in PnPDM~\cite{wu2024principled}.
The sampling of $\thetabm$ follows an analogous procedure. In the $\mathsf{LikelihoodStep}_\thetabm$, we sample:
\begin{equation*}
    \vbm^{(k)} \sim \exp\Big(-\tfrac{1}{2}\norm{\ybm - \Abm(\thetabm^{(k)})\ \vbm}_2^2 - \tfrac{1}{2\rho_\theta^2}\norm{\thetabm^{(k)} - \vbm}_2^2\Big),
\end{equation*}
where $\vbm \in \R^b$ is an auxiliary variable for the measurement operator parameters, and $\rho_\theta > 0$ controls the coupling between $\thetabm$ and $\vbm$.
In the $\mathsf{PriorStep}_\thetabm$, we sample:
\begin{equation}
    \thetabm^{(k+1)} \sim \exp\Big(-h(\thetabm) - \tfrac{1}{2\rho_\theta^2}\norm{\thetabm - \vbm^{(k)}}_2^2\Big).
\end{equation}
This step is also implemented using a DM $D_\beta$. To accelerate the Markov chain’s mixing time and avoid convergence to a bad local minima, we implement $k$-dependent annealing schedules for both $\rho_x$ and $\rho_\theta$.

\begin{table*}[htbp]
\caption{Quantitative evaluation of the proposed method in blind image deblurring. We highlighted the {\color{lightred}\textbf{best}} and {\color{lightblue}\underline{second best}} results, respectively. The \textit{Calibration} column highlights methods specifically designed to solve the blind inverse problem. The \textit{DM} column highlights methods that use diffusion models as priors. Noted how Blind-PnPDM outperforms several baseline methods.}
\begin{center}
\renewcommand\arraystretch{1.35}
\setlength{\tabcolsep}{2pt}
\begin{tabular}{lccccccccc}
\toprule
\multirow{2}{*}{\bf Method}  & \multirow{2}{*}{\it Calibration (Y/N)} & \multirow{2}{*}{\it DM (Y/N)} & \multicolumn{3}{c}{\textbf{Gaussian}} & & \multicolumn{3}{c}{\textbf{Motion}} \\
\cmidrule{4-6} \cmidrule{8-10} 
 & & & PSNR $\uparrow$ & SSIM $\uparrow$ & LPIPS $\downarrow$ &\ & PSNR $\uparrow$ & SSIM $\uparrow$ & LPIPS $\downarrow$ \\
\hline
 Pan-DCP & \cmark & \xmark & 21.50 & 0.685 & 0.408 && 17.07 & 0.514 & 0.507 \\
 DeblurGANv2 & \xmark & \xmark & 23.41 & 0.683 & 0.387 && 19.75 & 0.572 & 0.440 \\
 BlindDPS & \cmark & \cmark & 25.25 & 0.728 & {\color{lightblue}\underline{0.223}} && 21.95 & 0.638 & 0.288 \\
 GibbsDDRM &\cmark & \cmark & {\color{lightblue}\underline{26.49}} & {\color{lightblue}\underline{0.771}} & 0.240 && {\color{lightblue}\underline{25.94}} & {\color{lightblue}\underline{0.769}} & {\color{lightblue}\underline{0.204}} \\
 Ours & \cmark & \cmark & {\color{lightred}\textbf{27.13}} & {\color{lightred}\textbf{0.802}} & {\color{lightred}\textbf{0.180}} && {\color{lightred}\textbf{27.42}} & {\color{lightred}\textbf{0.795}} & {\color{lightred}\textbf{0.176}} \\
\bottomrule
\end{tabular}
\label{tab1}
\end{center}
\end{table*}
\section{Numerical Validation}
We validated Blind-PnPDM on blind image deblurring. The measurement operator in blind image deblurring can be modeled as $\Abm(\thetabm)\xbm = \thetabm * \xbm$, where $\thetabm$ is the unknown blur kernel, $\xbm$ is the unknown image, and $*$ is the convolution. 
We randomly selected 100 testing ground truth image from FFHQ~\cite{karras2019style} dataset. 
We used a pre-trained image diffusion models obtained from~\cite{chung2022come}. We followed~\cite{chung2023parallel} to train the kernel diffusion model on 100k generated blur kernel of size $64\times 64$ (both Gaussian and motion) for 5M steps with a small U-Net~\cite{dhariwal2021diffusion}.
We followed~\cite{chung2023parallel} to generate Gaussian and motion kernel for testing.
In our experiments, we set $K=30$, $\rho_x^{(k)}=\max(0.9^k\times0.3, 0.1)$ and $\rho_\theta^{(k)}=\max(0.9^k\times0.1, 0.05)$.

We compared Blind-PnPDM against several baseline methods, including Pan-DCP~\cite{pan2017deblurring},  DeblurGAN~\cite{kupyn2019deblurgan}, BlindDPS~\cite{chung2023parallel} and GibbsDDRM~\cite{murata2023gibbsddrm}. Pan-DCP is an optimization-based method that jointly estimates image and blur kernel. DeblurGAN is a supervised learning-based method that lacks the capability for kernel estimation, but can reconstruct images via direct inference. The results of DeblurGAN are obtained by running the published code with the pre-trained weights. BlindDPS and GibbsDDRM are diffusion model-based methods that jointly generate both the image and the kernel during their reverse diffusion processes.

Figure~\ref{fig:motion} and Figure~\ref{fig:gaussian} illustrate the reconstruction results using motion and Gaussian kernels, respectively. Both figures demonstrate that Blind-PnPDM can recover fine details of the human face, whereas non-DM methods produce smoother reconstructions. Moreover, Blind-PnPDM yields reconstruction results that are more consistent with the ground truth image and kernel, while other DM-based methods produce sharp but less consistent results. 
Table~\ref{tab1} presents a quantitative evaluation of the reconstruction performance, showing that Blind-PnPDM outperforms the baseline methods in both distortion-based metrics (PSNR/SSIM) and perceptual-based metrics (LPIPS).

\section{Conclusion}
In this paper, we present Blind-PnPDM as the first PnP sampling method for blind inverse problems. The key idea of Blind-PnPDM is to alternately sample the image and the unknown parameters of the measurement operator from their respective marginal distributions of the joint posterior. Each marginal distribution is further decomposed into sampling steps from the likelihood and prior distributions, with the prior modeled using diffusion models. 
We validate the effectiveness of Blind-PnPDM on the blind image deblurring problem. The results demonstrate that Blind-PnPDM achieves superior performance compared to both DM-based and non-DM-based methods, in terms of both quantitative metrics and qualitative visual quality.

\section{Acknowledgments}
Research presented in this article was supported by the NSF CAREER award CCF-2043134.

\end{document}